\documentclass[11pt,english]{article}

\usepackage[sc]{mathpazo}
\usepackage{geometry}
\usepackage{color}
\usepackage{array}
\usepackage{bbding}
\usepackage{multirow}
\usepackage[fleqn]{amsmath}
\usepackage{amssymb}
\usepackage{amsthm}
\usepackage{graphicx}
\usepackage{natbib}
\usepackage{lscape}
\usepackage[hidelinks]{hyperref}
\usepackage{comment}
\usepackage{bm}
\usepackage[title]{appendix}
\usepackage{longtable}
\usepackage{mathtools}
\usepackage{chngcntr}
\usepackage{authblk}
\usepackage{siunitx}
\usepackage{babel}
\usepackage{dsfont}
\usepackage[most]{tcolorbox}
\usepackage[short]{optidef}
\usepackage{cleveref}
\usepackage{booktabs}         
\usepackage{threeparttable}
\usepackage{makecell}
\usepackage{anyfontsize}
\usepackage{subcaption}
\usepackage{booktabs}
\usepackage{longtable}
\usepackage{float}
\usepackage{lineno}

\newcolumntype{C}[1]{>{\centering\arraybackslash}p{#1}}
\newcolumntype{A}{C{1.5cm}}   
\newcolumntype{B}{C{2.5cm}}   
\newcolumntype{D}{C{\dimexpr
  \linewidth - 4cm          
           - 8\tabcolsep    
           - 3\arrayrulewidth 
  \relax}}

\date{July 31, 2025}

\begin{document}
\title{Valuing Time in Silicon: Can Large Language Models Replicate Human Value of Travel Time}
	\author[1]{Yingnan Yan}
    \author[1]{Tianming Liu\footnote{Corresponding author. E-mail address: \textcolor{blue}{tianmliu@umich.edu} (T. Liu).}}
	\author[1]{Yafeng Yin}

	\affil[1]{\small\emph{Department of Civil and Environmental Engineering, University of Michigan, Ann Arbor, United States}\normalsize}
	\date{\today}
	\maketitle

\section*{Abstract}

\hspace{1.5em}As a key advancement in artificial intelligence, large language models (LLMs) are set to transform transportation systems. While LLMs offer the potential to simulate human travelers in future mixed-autonomy transportation systems, their behavioral fidelity in complex scenarios remains largely unconfirmed by existing research. This study addresses this gap by conducting a comprehensive analysis of the value of travel time (VOT) of three popular LLMs. We employ a full factorial experimental design to systematically examine LLMs' sensitivities to various transportation contexts, including the choice setting, travel purpose, and socio-demographic factors. Our results reveal a high degree of behavioral similarity between LLMs and humans. Some LLMs exhibit an aggregate VOT similar to that of humans, and all tested models demonstrate human-like sensitivity to travel purpose, income, and the time-cost trade-off ratios of the alternatives. Furthermore, the behavioral patterns of LLMs are highly consistent across varied contexts. However, while the behavior of every single model is highly robust, we also find some heterogeneity across models regarding the magnitude and direction of sensitivity to travel contexts and income elasticity. Overall, this study provides a foundational benchmark for the future development of LLMs as proxies for human travelers, demonstrating their robust decision-making capabilities while cautioning that misaligned magnitudes of economic trade-offs between humans and LLMs necessitate rigorous validation and additional conditioning of LLMs before their application.

\hfill\break%
\noindent\textit{Keywords} - Large language model, Value of travel time, Travel behavior, AI behavior \normalsize 

\newpage

\section{Introduction}
\hspace{1.5em}The advancement of artificial intelligence (AI) technology is continuously transforming the design, dynamics, and operations of transportation systems. Emerging technologies such as intelligent analytics, smart recommendation systems, and autonomous vehicles are constantly prompting the co-evolution of human and AI, and dramatically reshaping the future of transportation systems. 
Among the most recent advancements, the advent of large language models (LLMs) such as GPT-4 \citep{GPT4}, Gemini \citep{gemini}, and DeepSeek \citep{deepseekv3} represents a paradigm shift in AI. These models stand out in several aspects. First, built upon the transformer architecture \citep{attention} and guided by scaling laws \citep{kaplan2020scaling}, these models possess billions of parameters, enabling them to capture complex relationships and patterns with high fidelity and generalizability. Second, trained on massive corpora of textual data such as books, academic literature, and vast portions of the internet, LLMs possess the ability to generate human-like outputs and encode a broad spectrum of human knowledge with unprecedented quality. Finally, many of these models can accept multimodal and unstructured inputs, such as text, image, and video, making them exhibit impressive intelligence, interactivity, and high versatility for a wide range of applications. 
Building on the promise of earlier deep learning models for transportation, these modern LLMs are now considered tools with even greater transformative potential \citep{liu2025toward,nie2025exploring,yan2025large,karim2025large,wandelt2024large,mahmud2025integrating,xue2025transforming}.

Among the compelling applications of LLMs, a key direction is to utilize LLMs as simulated entities of human travelers in human-AI collaborative transportation systems. This suitability stems from the deep architecture and extensive training data, which allow LLMs to learn the patterns of human-generated data, mimic the behavior of diverse population groups, and generate responses and content that achieve unprecedented resemblance to humans \citep{horton2023}. Furthermore, LLMs exhibit in-context learning \citep{brown2020language} and contextual reasoning abilities, enabling them to adapt their responses and decisions to specific situations and instructions. These capabilities have been widely recognized and applied by researchers in various fields related to human behavior. For instance, LLMs have been leveraged to make human-like decisions in economic games, where they adapt strategies to assigned roles \citep{horton2023}. Similarly, they exhibit human-like characteristics in psychological experiments \citep{aher2023using} and serve as participants in survey research to enhance design quality \citep{jansen2023employing}. However, the behavioral evidence from these domains does not guarantee validity in transportation contexts, where decision-making is uniquely constrained by spatial dynamics, mode availability, and the specific trade-off between time and monetary cost. Consequently, recent work has begun testing LLMs in specific transportation tasks, such as mode and route choice \citep{wang2024ai}, activity scheduling \citep{liu2025toward}, and critical decision-making in autonomous driving \citep{huang2024drivlme}.

As simulated entities of human travelers, LLMs may significantly impact the dynamics and design of future transportation systems in two critical ways. First, they can power autonomous agents that will coexist with humans, and the behavior of these agents will directly reshape complex system dynamics like traffic flow and mode choice. Second, their human-like capabilities allow them to generate rich synthetic travel data, overcoming longstanding data collection hurdles for researchers and providing a powerful new resource for analysis \citep{zhang2024agentic, li2024more}. This data is especially useful for tasks such as hypothesis generation, policy evaluation, and large-scale agent-based modeling \citep{liu2025toward}. Given these transformative roles, a thorough understanding of LLM behavior is essential to guide their future integration into transportation safely and effectively.

To safely integrate LLMs as autonomous agents in joint human-AI transportation systems, we must first analyze the values guiding their decisions and evaluate their alignment with those of humans. However, with few exceptions \citep{xu2025morality}, current research into LLM behavior in transportation has focused primarily on replicating the final choices of human travelers. As a result, the underlying values that guide these models' decisions in traveler contexts, and the degree to which they align with humans, remain largely unexplored. Furthermore, beyond transportation, studies on LLM behaviors in cognitive science, social science, and other engineering domains show mixed results regarding their alignment with human behavior and decisions. While LLMs demonstrate remarkable similarities to humans in some areas, including exhibiting similar economic behavior \citep{horton2023,aher2023using}, possessing cohesive ``personalities'' \citep{ji2024persona, chuangalign}, and even passing the Turing Test \citep{mei2024turing}, they also show significant deviations. For instance, LLMs can display higher rationality than humans \citep{econrationGPT}, exhibit systematic biases \citep{rozado2023political, taubenfeld2024systematic}, and inadequately represent certain population groups \citep{santurkar2023whose,wang2025large}. Additionally, the existing studies also overlooked some important properties that could be critical for transportation, such as LLMs' sensitivity to travel contexts and the trade-off ratios among available alternatives. In summary, existing evidence is insufficient to confirm whether LLMs can reliably simulate human behavior in transportation scenarios, necessitating further targeted research.

To fill this research gap, in this paper, we investigate the behavior of three popular LLMs: GPT-4o, Gemini-2.5-pro, and Claude-Sonnet-4 through the lens of value of travel time (VOT) to understand their values and human resemblance. We focus on VOT for two reasons: First, VOT is the fundamental metric governing cost-benefit analysis; therefore,  meaningful alignment in this metric is a prerequisite for the LLMs to function as simulated entities of human travelers and engage in agent-based simulations and synthetic data generation. Second, its nature as a time-cost trade-off provides a rigorous test of the model's responsiveness to the context. Methodologically, we use factorial design and prompt engineering based on a human VOT study \citep{calfee2001} to quantify the VOT of the tested LLMs and its sensitivity to transportation contexts. Furthermore, we compare the VOT of LLMs and humans both qualitatively and quantitatively to gain insight into their similarities and differences. Our results indicate a high degree of behavioral similarity between the LLMs and humans on VOT, and demonstrate that the LLMs exhibits significant human-like sensitivity to transportation contexts. Additionally, our analysis reveals that the LLMs exhibits strong internal consistency across different travel contexts. However, we also discover that the characteristics of different models exhibit clear discrepancies, where the travel context sensitivity of some LLMs is not as pronounced as that of human travelers, particularly in the income elasticity of VOT. To our knowledge, this is the first study to systematically analyze LLMs' valuation of travel time. Our findings offer new insights into the internal value of LLMs and their alignment with humans in terms of travel simulation, thereby contributing to a fundamental understanding of LLM behaviors in transportation and offering new practical implications for their future applications.

The remainder of this paper proceeds as follows. \Cref{sec:LR} first reviews the existing literature on LLM behavior, establishing the current state of knowledge and identifying the key research gaps this study addresses. \Cref{sec:method} then details our experimental methodology, including the full factorial design, the prompting framework, and the analytical models used to quantify the VOT of LLMs. \Cref{sec:results} presents and discusses the core findings from our analysis. Finally, \Cref{sec:conclusion} concludes by summarizing the study's contributions, discussing their implications, and outlining promising avenues for future research.

\section{Literature Review} \label{sec:LR}
\hspace{1.5em}In transportation, research on LLM decision-making currently focuses on evaluating LLM's synthetic travel data generation accuracy on tasks such as travel choices, activity generation, and trajectory prediction. \cite{wang2023would}, \cite{beneduce2025large}, and \cite{feng2024agentmove} showed that LLMs can effectively simulate travelers' destinations by choosing suitable destinations based on travel history and traveler profile. \cite{li2024more} illustrated that LLMs can generate realistic travel activities and diaries by diagnosing travelers' needs and intentions from their profiles. \cite{jiawei2024large} and \cite{ge2025llm} leveraged LLMs to predict the travel trajectories of travelers based on their travel history. They showed that LLMs can replicate personalized travel patterns with a satisfactory degree of accuracy and highlighted the interpretability gained from semantic data. \cite{mo2023large}, \cite{liu2024can}, \cite{alsaleh2025towards}, and \cite{liu2025aligning} conducted experiments on LLMs' travel mode choices and compared the results with human choices. Collectively, they discovered detailed discrepancies between LLMs and humans in the proportion of choices for each mode, and the second of which proposed multiple methods that significantly improved the fidelity of LLMs' choices to human behavior. However, existing research has mainly focused on whether LLMs replicate the travel choices and outcomes of humans in specific scenarios. Consequently, whether this alignment is merely surface-level or if these models operate on the same underlying values that guide human travelers' decision-making remains an unexplored critical question. Regarding the LLMs' internal value in transportation scenarios, currently, the sole discussion is provided by \cite{xu2025morality}. They designed an ethical dilemma for autonomous vehicles, where LLMs had to choose between saving passengers or pedestrians. They showed that LLMs exhibited a tendency to protect the vulnerable or valuable group of people rather than choosing the group with a higher survival probability, which is a commonly observed but not universal case for humans. In summary, while existing research demonstrates the potential of LLMs in transportation, it offers only a fragmented understanding of their behavior, leaving their underlying values largely unexplored. Further assessment is therefore necessary to move beyond replicating outcomes and toward a fundamental understanding of how these models make decisions in critical transportation contexts.

Given the well-established economic foundation of travel behavior, findings from studies of LLMs' economic behavior can serve as valuable references. These studies underscored LLMs' potential to generate human-like behaviors, yet they also revealed various instances where LLM behaviors diverged from human behaviors. In a pioneering study, \cite{horton2023} replicated classic behavioral economics experiments using GPT-3. The results demonstrated that LLMs can exhibit human-like decision-making, engage in basic role-playing, show sensitivity to how questions are worded, and exhibit human-like status quo bias. However, their specific choices still differed noticeably from those of humans. Subsequent studies expanded the scope of economic games and investigated newer LLM versions \citep{akata2025playing,xie2024can,fan2024can,aher2023using}, similarly finding that LLMs could comprehend game settings and adopt human-like strategies. Furthermore, studies have found that LLMs exhibit similar behavioral biases in decision-making as humans, such as risk aversion \citep{song2025can} and mental accounting \citep{leng2024can} in some economic contexts. However, LLMs frequently exhibited some principles that are not consistently observed in human economic behavior, such as strong reciprocity \citep{xie2024can, leng2023llm}, heightened fairness and cooperation \citep{brookins2024playing}, higher rationality \citep{econrationGPT}, and reaction to high-risk situations \citep{leng2024can}, irrespective of socio-demographic characteristics assigned to the models. While these studies revealed LLMs' considerable capability for human-like decisions, they also highlighted behavioral discrepancies between LLMs and humans.

Beyond economics, studies in psychology and sociology similarly reveal that while LLMs often emulate human social behavior, they also exhibit limitations and discrepancies. In psychology, \cite{miotto2022gpt} applied the HEXACO personality assessment to LLMs, finding that LLMs exhibit human-like personalities. Building on this, \cite{ji2024persona} demonstrated that LLMs can consistently manifest a personality based on instructions and socio-demographic information. Furthermore, \cite{chuangalign} showed that LLMs can successfully assume correct attitudes when provided with out-of-context attitude information. \cite{argyle2023out} utilized LLMs to simulate human survey participants, highlighting LLMs' excellent representativeness of a wide range of population groups. Their lexical responses concerning political perceptions passed the Turing Test, and their responses to quantitative survey questions reflected high resemblance to multiple population groups, suggesting high fidelity for using LLMs to understand real human perceptions. \cite{mei2024turing} also discovered LLM survey respondents passing the Turing Test, revealing the high human-resemblance of LLMs.
However, \cite{tjuatja2024llms} demonstrated that LLMs did not exhibit typical human response biases. Concurrently, \cite{chen2025manager} specified some human biases where LLMs exhibited the opposite behavior rather than replicating them. \cite{rozado2023political} and \cite{motoki2024more} uncovered biases of LLMs in their political opinions. 
Moreover, \cite{wang2025large}, \cite{lee2024large}, and \cite{dominguez2024questioning} revealed that the response diversity among LLMs assigned with different identities or demographic groups was less pronounced than that of humans, and LLM responses within the same group were more homogeneous. Similarly, research has found that LLMs also exhibit social identity bias \citep{hu2025generative,sun2023aligning,shin2024ask} and could not adequately represent the opinions of certain population groups \citep{santurkar2023whose,wang2025large}. 

While the existing literature presents varied findings on LLM behavior, critical gaps remain within the context of travel behavior research. First, the fundamental values guiding LLM decisions in the context of transportation are barely understood. LLMs' behavior regarding the value of travel time (VOT), a cornerstone of human travel behavior analysis, has not yet been studied. Furthermore, existing studies have not examined the consistency of LLMs' values across different contexts, which is a critical requirement for ensuring the robustness of their integration into transportation. Second, key aspects of the LLMs' decision-making process remain unexplored. Although prior work has shown that LLMs exhibit some human-like biases in decision-making \citep{horton2023,leng2024can,song2025can}, their responses to alternatives with varying trade-off ratios are yet to be studied. This is a crucial omission, as the trade-off ratios, especially between time and cost, heavily influence human travel choices and have significant implications for understanding traveler values and for planning applications \citep{fowkes1988,hess2018}. Addressing these gaps is essential to ensure that LLMs align not only with human choice outcomes but also with their underlying values and decision-making tendencies, which this study aims to investigate.

\section{Methodology} \label{sec:method}
\begin{figure}[h!]
\centering
\includegraphics[width=1\textwidth]{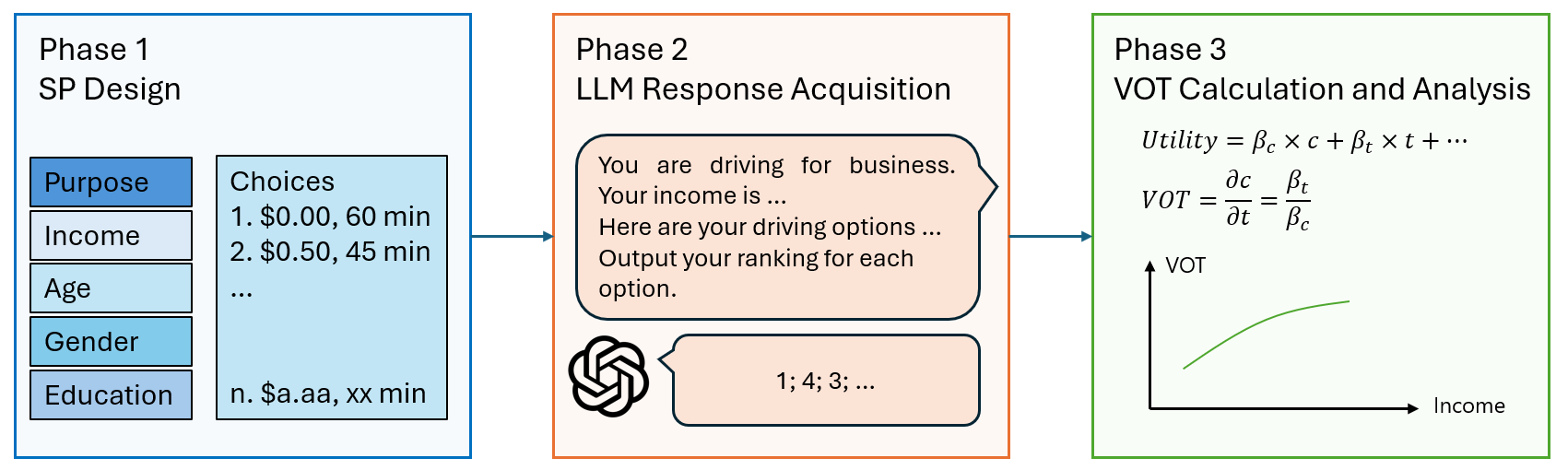}
\caption{The three-phase method to estimate the VOT of LLMs}
\label{fig:method}
\end{figure}

\hspace{1.5em}To systematically investigate and measure the VOT of LLMs, we apply the classical stated preference (SP) survey method and treat the LLMs as survey respondents. Furthermore, to account for contextual impacts on VOT and test the sensitivity of LLM VOT in different contexts, we adopt the full factorial design method. As illustrated in \Cref{fig:method}, our method includes three main phases: (1) designing a full factorial for the SP survey context and questions; (2) collecting response data by prompting an LLM with the designed SP survey; and (3) calculating the VOT using statistical models and comparing the results with established human values. The three phases are elaborated in \Cref{subsec:design}, \Cref{subsec:data}, and \Cref{subsec:vot}, respectively.

\subsection{Factorial design of SP contexts} \label{subsec:design}
\hspace{1.5em}First, we specify the choice settings and contexts. To ensure our LLM estimations are comparable with human data, we adopt the SP choice design from a human VOT study by \cite{calfee2001}. In this design, respondents who travel by car are asked to rank 13 travel alternatives based on their preferences, where each alternative is defined by its travel time, cost (in the form of toll), and truck presence. The specific alternatives of \cite{calfee2001} are detailed in choice package 1 of \Cref{tab:allchoices}. The original study included two distinct but related sets of choices: the second set has costs and times scaled up by a factor of 1.5, which preserves the cost-time ratios of corresponding alternatives. In our study, we utilize both choice sets. For clarity, we use ``choice set'' to refer to the set of 13 alternatives and use ``choice package'' to refer to an entire design that includes two choice sets throughout this paper.

\begin{longtable}{|A|B|D|}
\hline
\makecell[l]{\textbf{Choice}\\\textbf{Package}} & \makecell[l]{\textbf{Average trade-}\\\textbf{off ratio}} & \textbf{Alternatives}\\\hline
\endfirsthead
\multicolumn{3}{@{}l}{\textit{Table \thetable{} – continued}}\\\hline
\makecell[l]{\textbf{Choice}\\\textbf{Package}} &
\makecell[l]{\textbf{Average trade-}\\\textbf{off ratio}} &
\textbf{Alternatives}\\\hline
\endhead
\hline
\multicolumn{3}{r@{}}{\textit{continued on next page}}\\
\endfoot
\endlastfoot
%
1 & \$6.6/h &
\begin{minipage}[t]{\linewidth}
  \begingroup
    \setlength{\tabcolsep}{4pt}
    \scriptsize
    \resizebox{\linewidth}{!}{%
  \begin{tabular}{@{}c c c c | c c c c@{}}
  \multicolumn{4}{c|}{\textbf{Choice set 1}} &
  \multicolumn{4}{c}{\textbf{Choice set 2}} \\\hline
  \textbf{Alt.}&\textbf{Cost \$}&\textbf{Time min}&\textbf{Trucks}&
  \textbf{Alt.}&\textbf{Cost \$}&\textbf{Time min}&\textbf{Trucks}\\\hline
  1 & 0.00 & 40 & Yes & 1  & 0.00 & 60 & Yes \\
  2 & 0.35 & 30 & Yes & 2  & 0.50 & 45 & Yes \\
  3 & 0.70 & 30 & Yes & 3  & 1.00 & 45 & Yes \\
  4 & 1.00 & 30 & Yes & 4  & 1.50 & 45 & Yes \\
  5 & 0.35 & 20 & Yes & 5  & 0.50 & 30 & Yes \\
  6 & 0.70 & 20 & Yes & 6  & 1.00 & 30 & Yes \\
  7 & 1.35 & 20 & Yes & 7  & 2.00 & 30 & Yes \\
  8 & 2.00 & 10 & Yes & 8  & 3.00 & 15 & Yes \\
  9 & 3.35 & 10 & Yes & 9  & 5.00 & 15 & Yes \\
 10 & 0.35 & 40 & No  &10  & 0.50 & 60 & No  \\
 11 & 0.70 & 30 & No  &11  & 1.00 & 45 & No  \\
 12 & 1.75 & 20 & No  &12  & 2.50 & 30 & No  \\
 13 & 4.00 & 10 & No  &13  & 6.00 & 15 & No  \\
  \end{tabular}}%
\endgroup
\end{minipage}\\ \hline
%
2 & \$9.7/h &
\begin{minipage}[t]{\linewidth}
  \centering\scriptsize\setlength{\tabcolsep}{3pt}%
  \resizebox{\linewidth}{!}{%
  \begin{tabular}{@{}c c c c | c c c c@{}}
  \multicolumn{4}{c|}{\textbf{Choice set 1}} &
  \multicolumn{4}{c}{\textbf{Choice set 2}} \\\hline
  \textbf{Alt.}&\textbf{Cost \$}&\textbf{Time min}&\textbf{Trucks}&
  \textbf{Alt.}&\textbf{Cost \$}&\textbf{Time min}&\textbf{Trucks}\\\hline
  1 & 0.00 & 40 & Yes & 1 & 0.00 & 60 & Yes \\
  2 & 0.50 & 30 & Yes & 2 & 0.75 & 45 & Yes \\
  3 & 1.00 & 30 & Yes & 3 & 1.50 & 45 & Yes \\
  4 & 1.50 & 30 & Yes & 4 & 2.25 & 45 & Yes \\
  5 & 0.50 & 20 & Yes & 5 & 0.75 & 30 & Yes \\
  6 & 1.00 & 20 & Yes & 6 & 1.50 & 30 & Yes \\
  7 & 2.00 & 20 & Yes & 7 & 3.00 & 30 & Yes \\
  8 & 3.00 & 10 & Yes & 8 & 4.50 & 15 & Yes \\
  9 & 5.00 & 10 & Yes & 9 & 7.50 & 15 & Yes \\
 10 & 0.50 & 40 & No  &10 & 0.75 & 60 & No  \\
 11 & 1.00 & 30 & No  &11 & 1.50 & 45 & No  \\
 12 & 2.70 & 20 & No  &12 & 4.00 & 30 & No  \\
 13 & 6.00 & 10 & No  &13 & 9.00 & 15 & No  \\
  \end{tabular}}%
\end{minipage}\\ \hline
%
3 & \$19.4/h &
\begin{minipage}[t]{\linewidth}
  \centering\scriptsize\setlength{\tabcolsep}{3pt}%
  \resizebox{\linewidth}{!}{%
  \begin{tabular}{@{}c c c c | c c c c@{}}
  \multicolumn{4}{c|}{\textbf{Choice set 1}} &
  \multicolumn{4}{c}{\textbf{Choice set 2}} \\\hline
  \textbf{Alt.}&\textbf{Cost \$}&\textbf{Time min}&\textbf{Trucks}&
  \textbf{Alt.}&\textbf{Cost \$}&\textbf{Time min}&\textbf{Trucks}\\\hline
  1 & 0.00  & 40 & Yes & 1 & 0.00  & 60 & Yes \\
  2 & 1.00  & 30 & Yes & 2 & 1.50  & 45 & Yes \\
  3 & 2.00  & 30 & Yes & 3 & 3.00  & 45 & Yes \\
  4 & 3.00  & 30 & Yes & 4 & 4.50  & 45 & Yes \\
  5 & 1.00  & 20 & Yes & 5 & 1.50  & 30 & Yes \\
  6 & 2.00  & 20 & Yes & 6 & 3.00  & 30 & Yes \\
  7 & 4.00  & 20 & Yes & 7 & 6.00  & 30 & Yes \\
  8 & 6.00  & 10 & Yes & 8 & 9.00  & 15 & Yes \\
  9 &10.00  & 10 & Yes & 9 &15.00  & 15 & Yes \\
 10 & 1.00  & 40 & No  &10 & 1.50  & 60 & No  \\
 11 & 2.00  & 30 & No  &11 & 3.00  & 45 & No  \\
 12 & 5.35  & 20 & No  &12 & 8.00  & 30 & No  \\
 13 &12.00  & 10 & No  &13 &18.00  & 15 & No  \\
  \end{tabular}}%
\end{minipage}\\ \hline
%
4 & \$29.1/h &
\begin{minipage}[t]{\linewidth}
  \centering\scriptsize\setlength{\tabcolsep}{3pt}%
  \resizebox{\linewidth}{!}{%
  \begin{tabular}{@{}c c c c | c c c c@{}}
  \multicolumn{4}{c|}{\textbf{Choice set 1}} &
  \multicolumn{4}{c}{\textbf{Choice set 2}} \\\hline
  \textbf{Alt.}&\textbf{Cost \$}&\textbf{Time min}&\textbf{Trucks}&
  \textbf{Alt.}&\textbf{Cost \$}&\textbf{Time min}&\textbf{Trucks}\\\hline
  1 & 0.00 & 40 & Yes & 1 & 0.00 & 60 & Yes \\
  2 & 1.50 & 30 & Yes & 2 & 2.25 & 45 & Yes \\
  3 & 3.00 & 30 & Yes & 3 & 4.50 & 45 & Yes \\
  4 & 4.50 & 30 & Yes & 4 & 6.75 & 45 & Yes \\
  5 & 1.50 & 20 & Yes & 5 & 2.25 & 30 & Yes \\
  6 & 3.00 & 20 & Yes & 6 & 4.50 & 30 & Yes \\
  7 & 6.00 & 20 & Yes & 7 & 9.00 & 30 & Yes \\
  8 & 9.00 & 10 & Yes & 8 &13.50 & 15 & Yes \\
  9 &15.00 & 10 & Yes & 9 &22.50 & 15 & Yes \\
 10 & 1.50 & 40 & No  &10 & 2.25 & 60 & No  \\
 11 & 3.00 & 30 & No  &11 & 4.50 & 45 & No  \\
 12 & 8.00 & 20 & No  &12 &12.00 & 30 & No  \\
 13 &18.00 & 10 & No  &13 &27.00 & 15 & No  \\
  \end{tabular}}%
\end{minipage}\\ \hline
%
5 & \$38.8/h &
\begin{minipage}[t]{\linewidth}
  \centering\scriptsize\setlength{\tabcolsep}{3pt}%
  \resizebox{\linewidth}{!}{%
  \begin{tabular}{@{}c c c c | c c c c@{}}
  \multicolumn{4}{c|}{\textbf{Choice set 1}} &
  \multicolumn{4}{c}{\textbf{Choice set 2}} \\\hline
  \textbf{Alt.}&\textbf{Cost \$}&\textbf{Time min}&\textbf{Trucks}&
  \textbf{Alt.}&\textbf{Cost \$}&\textbf{Time min}&\textbf{Trucks}\\\hline
  1 & 0.00 & 40 & Yes & 1 & 0.00  & 60 & Yes \\
  2 & 2.00 & 30 & Yes & 2 & 3.00  & 45 & Yes \\
  3 & 4.00 & 30 & Yes & 3 & 6.00  & 45 & Yes \\
  4 & 6.00 & 30 & Yes & 4 & 9.00  & 45 & Yes \\
  5 & 2.00 & 20 & Yes & 5 & 3.00  & 30 & Yes \\
  6 & 4.00 & 20 & Yes & 6 & 6.00  & 30 & Yes \\
  7 & 8.00 & 20 & Yes & 7 &12.00  & 30 & Yes \\
  8 &12.00 & 10 & Yes & 8 &18.00  & 15 & Yes \\
  9 &20.00 & 10 & Yes & 9 &30.00  & 15 & Yes \\
 10 & 2.00 & 40 & No  &10 & 3.00  & 60 & No  \\
 11 & 4.00 & 30 & No  &11 & 6.00  & 45 & No  \\
 12 &10.70 & 20 & No  &12 &16.00  & 30 & No  \\
 13 &24.00 & 10 & No  &13 &36.00  & 15 & No  \\
  \end{tabular}}%
\end{minipage}\\ \hline
%
6 & \$48.5/h &
\begin{minipage}[t]{\linewidth}
  \centering\scriptsize\setlength{\tabcolsep}{3pt}%
  \resizebox{\linewidth}{!}{%
  \begin{tabular}{@{}c c c c | c c c c@{}}
  \multicolumn{4}{c|}{\textbf{Choice set 1}} &
  \multicolumn{4}{c}{\textbf{Choice set 2}} \\\hline
  \textbf{Alt.}&\textbf{Cost \$}&\textbf{Time min}&\textbf{Trucks}&
  \textbf{Alt.}&\textbf{Cost \$}&\textbf{Time min}&\textbf{Trucks}\\\hline
  1 & 0.00 & 40 & Yes & 1 & 0.00 & 60 & Yes \\
  2 & 2.50 & 30 & Yes & 2 & 3.75 & 45 & Yes \\
  3 & 5.00 & 30 & Yes & 3 & 7.50 & 45 & Yes \\
  4 & 7.50 & 30 & Yes & 4 &11.25 & 45 & Yes \\
  5 & 2.50 & 20 & Yes & 5 & 3.75 & 30 & Yes \\
  6 & 5.00 & 20 & Yes & 6 & 7.50 & 30 & Yes \\
  7 &10.00 & 20 & Yes & 7 &15.00 & 30 & Yes \\
  8 &15.00 & 10 & Yes & 8 &22.50 & 15 & Yes \\
  9 &25.00 & 10 & Yes & 9 &37.50 & 15 & Yes \\
 10 & 2.50 & 40 & No  &10 & 3.75 & 60 & No  \\
 11 & 5.00 & 30 & No  &11 & 7.50 & 45 & No  \\
 12 &13.35 & 20 & No  &12 &20.00 & 30 & No  \\
 13 &30.00 & 10 & No  &13 &45.00 & 15 & No  \\
  \end{tabular}}%
\end{minipage}\\ \hline
\caption{All choice packages used in this study}\label{tab:allchoices}
\end{longtable}


In addition to the choice alternatives, to effectively simulate a human traveler's perspective with the LLMs, we also specify its socio-demographic identity and the travel context, both of which are shown to be correlated with human travelers' VOT \citep{meta2009,axhausen2004swiss,gunn2001,USDOT2016VTT}. To capture their impacts, we incorporate six factors in our experiment design: the choice setting, travel purpose, and four socio-demographic aspects. The socio-demographic aspects, income, gender, age, and highest educational background are selected due to their high relevance to the VOT and their common inclusion in survey data. Consequently, we employ a full factorial design with these factors to systematically investigate the LLMs' VOT value and sensitivity in different contexts. The specific levels of the factors included in this design are presented below in \Cref{tab:factor}. Overall, this full factorial design results in 768 combinations, providing cases for a variety of personal identities and travel contexts to study LLM's behavioral sensitivity.

{
\renewcommand{\arraystretch}{1.1}
\begin{table}[h!]
\centering
\begin{tabular}{cll}
\hline
  & Attributes & Levels \\
\hline
1 & Average trade-off ratio of choice package & 6.6, 9.7, 19.4, 29.1, 38.8, 48.5 (USD/h) \\
2 & Travel purpose & Personal, Business, Commute, Leisure \\
3 & Income & 15, 25, 35, 50 (USD/h) \\
4 & Gender & Male, Female \\
5 & Age & 20s, 50s \\
6 & Education level & High school, Bachelor's degree or higher \\
\hline
\end{tabular}
\caption{Levels of the full factorial design for the SP setting}
\label{tab:factor}
\end{table}
}

For socio-demographic factors, we set the levels according to the 2023 U.S. Census and related literature of human VOT. On travel purpose, we set four different levels ranging from leisure travel to business travel according to existing VOT meta-analyses \citep{meta2009,binsuwadan2023income} to cover a wide range of factors and facilitate comparison with human benchmarks. The income levels are set according to the 2023 historical income tables of households issued by the U.S. Census Bureau \citep{CensusBureau2023_HistIncomeHH}. The education levels' descriptions are also selected according to the standard of the U.S. Census. For age, we establish two distinct levels representing younger and older travelers, two group chosen for comparison due to the existing research on their travel behavior \citep{malokin2021millennials}. For the choice setting, we differentiate the levels by selecting multiple choice packages with six different trade-off ratios. In SP surveys, the trade-off ratio, defined as the absolute ratio of the cost difference to the time difference between alternatives, has been shown to affect the results \citep{fowkes1988,hess2018}. Therefore, in this factorial design, we construct multiple choice packages with different average trade-off ratios to examine the sensitivity of LLM to the choice setting. We adjust the trade-off ratios of choice package 1 in \Cref{tab:allchoices} by rescaling the travel costs in the alternatives. Based on the typical express lane tolls in California (ranging from \$0.10/mile non-peak to \$0.35/mile peak) and the average U.S. commuting time (26.8 minutes in 2023 census \citep{census_commute2023}), we modify the travel costs in the choice sets accordingly and keep the travel times unchanged. The new choice package, presented in choice packages 4 in \Cref{tab:allchoices}, preserves the two multi-alternative choice task structures, with an average trade-off ratio of approximately \$29.1/h among alternatives. Next, to create further variations in trade-off ratio, we generate several additional choice packages with varying trade-off ratios by scaling the travel costs. Specifically, we multiply the costs by $\frac{1}{3}$, $\frac{2}{3}$, 1, $\frac{4}{3}$, and $\frac{5}{3}$. This yields five distinct average trade-off ratios: \$9.7/h, \$19.4/h, \$29.1/h, \$38.8/h, and \$48.5/h. These five choice settings are presented in choice packages 2, 3, 4, 5, and 6 in \Cref{tab:allchoices}, respectively, and together with the trade-off ratio in \cite{calfee2001}, form the six levels of trade-off ratios in our full factorial design.

\subsection{LLM Response Collection} \label{subsec:data}
\hspace{1.5em}In this phase, we design the prompt to collect the LLMs' survey response. For this study, we select three LLMs: OpenAI's GPT-4o, Google's Gemini-2.5-pro, and Anthropic's Claude-Sonnet-4. These three models are selected since they represent the current state-of-the-art ``general-purpose'' generative AI models, which are not only used as proxies for human behavior, but are also widely accessible for other research applications. 

\begin{figure}[h!]
\centering
\includegraphics[width=1\textwidth]{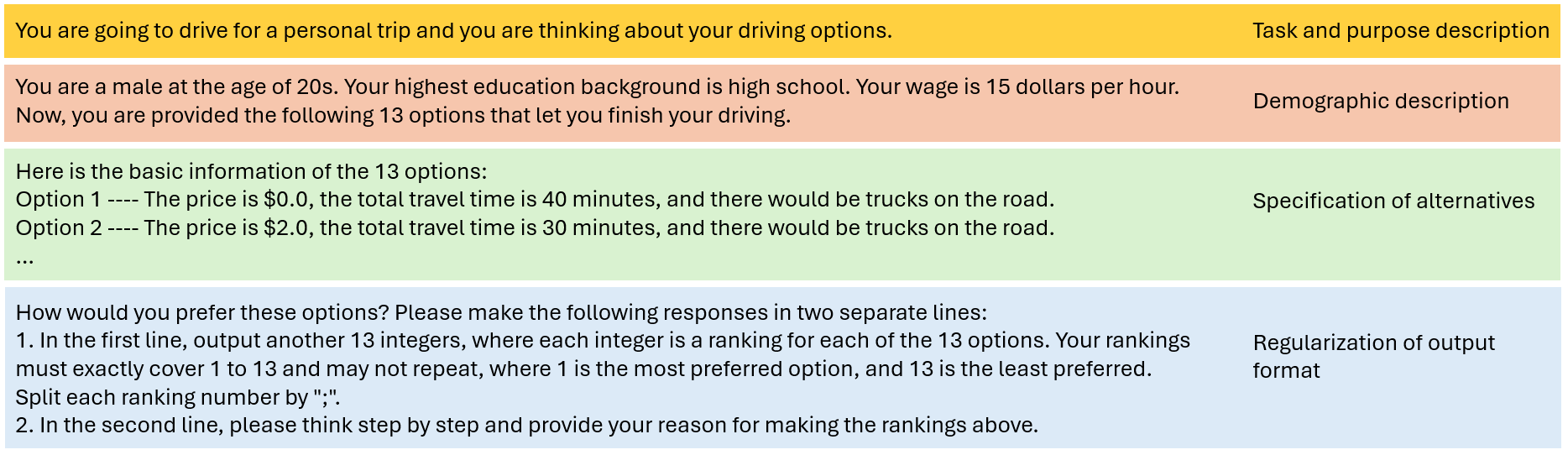}
\caption{Prompting framework for administering the SP survey to the LLMs.}
\label{fig:prompt}
\end{figure}

We employ a unified prompt setting across all LLMs to generate their responses in a normalized format. Consistent with existing studies in transportation \citep{liu2024can,xu2025morality}, the prompt first describes the context and assigns a sociodemographic profile to the LLMs, then specifies the SP alternatives, and finally instructs the LLMs to provide the rankings and standardizes the output format. In each response, consistent with the human survey \citep{calfee2001}, the LLMs are asked to provide its rankings of the 13 alternatives in the choice set, with rank 1 indicating the most preferred option and rank 13 the least preferred, ensuring that no ranks are repeated. We also require the LLMs to output the reason for the rankings to regulate output generation and reduce irrelevant or arbitrary responses \citep{liu2024can}\footnote{We note that asking the LLMs to articulate reasoning does not guarantee true cognitive alignment with human mental processes due to the complexity of current model architectures.}. An example prompt is presented in \Cref{fig:prompt}. We access all three LLMs through their respective APIs. To capture the full distribution of potential behaviors and introduce necessary stochasticity for generating diverse survey responses, we set the sampling temperature to 1 for all experiments.

We conduct the response collection in two steps. First, to have a baseline understanding of the VOT values of LLMs and to analyze the LLMs' general behavioral alignment and sensitivity to socio-demographic factors, we query all three models using Choice Package 1 in \Cref{tab:allchoices} (the baseline setting from \cite{calfee2001}). Second, to investigate sensitivity to the choice setting, specifically the trade-off ratios between time and cost among different available alternatives, we utilize GPT-4o as the representative model to complete the full factorial design across all six choice packages listed in \Cref{tab:factor}. We limit this extensive choice-setting analysis to a single model to manage computational costs. For each combination of factors, we collect 60 responses from every LLM for each of the two choice sets. This generates a total of 92,160 responses of ranked preferences for GPT-4o, and 15,360 responses for each of the other two LLMs, forming a comprehensive dataset for statistical analysis.

\subsection{VOT Calculation and Analysis} \label{subsec:vot}
\hspace{1.5em}After acquiring the responses from LLMs, we estimate their alternative preference ranking model and calculate the VOT. For the ranking model, we use the rank-ordered logit model \citep{rologit} to calibrate a linear utility function,
\begin{equation}
\begin{split}
    U(\boldsymbol{x})&=\boldsymbol{\beta}^T\boldsymbol{x}+\epsilon \\
    &=\beta_{cost}\times \texttt{cost}+\beta_{time}\times \texttt{time} + {\beta}_{truck}\times \texttt{truck}+\epsilon,    
\end{split}
\label{eq1}
\end{equation}
where $\boldsymbol{\beta}=[\beta_{cost}, \beta_{time}, {\beta}_{truck}]^T$ is the parameter vector, including the coefficients for the travel cost, travel time, and truck existence condition, respectively; $\boldsymbol{x}=[\texttt{cost}, \texttt{time}, \texttt{truck}]^T$ is the attribute vector of an alternative, including the travel cost $\texttt{cost}$, the travel time $\texttt{time}$, and the dummy variable indicating truck existence on the road, $\texttt{truck}$; $\epsilon$ is the error term, which follows the Type-I extreme value distribution. The utility function's specification is consistent with \cite{calfee2001} to facilitate comparison of results.

Denote each run in the full factorial design in \Cref{tab:factor} as $k=1,2,...,768$, and the samples obtained in each run by the corresponding two parallel packages in \Cref{tab:allchoices} as $i=1,2,...,120$. For each sample's ranking, by extending the multinomial logit model, we can obtain a closed-form solution for the probability of observing a certain order of ranking the 13 alternatives for the model specified in \Cref{eq1},
\begin{equation}    P[U(\boldsymbol{x_{i,(1)}^k})>U(\boldsymbol{x_{i,(2)}^k})>...>U(\boldsymbol{x_{i,(13)}^k})]=\prod_{h=1}^{12} \frac{e^{\boldsymbol{\beta}^T\boldsymbol{x_{i,(h)}^k}}}{\sum_{m=h}^{13}e^{\boldsymbol{\beta}^T\boldsymbol{x_{i,(m)}^k}}},
\label{eq2}
\end{equation}
where $\boldsymbol{x_{i,(1)}^k}, \boldsymbol{x_{i,(2)}^k}, ..., \boldsymbol{x_{i,(13)}^k}$ denote the attribute vectors for the alternatives ranked as 1, 2, ..., 13, and $U(\boldsymbol{x_{i,(1)}^k}), U(\boldsymbol{x_{i,(2)}^k}), ..., U(\boldsymbol{x_{i,(13)}^k})$ are their corresponding utilities, respectively.

Therefore, for run $k$, the log-likelihood function $L_k(\boldsymbol{\beta})$ can be expressed as,
\begin{equation}
    \begin{split}
        L_k(\boldsymbol{\beta})&=\sum_{i=1}^{N}\ln\left[\prod_{h=1}^{12} \frac{e^{\boldsymbol{\beta}^T\boldsymbol{x_{i,(h)}^k}}}{\sum_{m=h}^{13}e^{\boldsymbol{\beta}^T\boldsymbol{x_{i,(m)}^k}}}\right] \\
        &= \sum_{i=1}^{N}\sum_{h=1}^{12}\boldsymbol{\beta}^T\boldsymbol{x_{i,(h)}^k}-\sum_{i=1}^{N}\sum_{h=1}^{12}        \ln\left[\sum_{m=h}^{13}e^{\boldsymbol{\beta}^T\boldsymbol{x_{i,(m)}^k}}\right],
    \end{split}
\label{eq3}
\end{equation}
where $N=120$ is the sample size for every run. 

The estimated parameters $\hat{\boldsymbol{\beta_k}}=[\hat{\beta}_{cost}^k,\hat{\beta}_{time}^k,\hat{\beta}_{truck}^k]^T$ can be obtained by the solution of the following maximum likelihood estimation problem,
\begin{equation}
\begin{aligned}
  \underset{\boldsymbol{\beta_k}}{\text{max}}\;\; & L_k(\boldsymbol{\beta}) = \sum_{i=1}^{N}\sum_{h=1}^{12}\boldsymbol{\beta}^T\boldsymbol{x_{i,(h)}^k}-\sum_{i=1}^{N}\sum_{h=1}^{12} \ln\left[\sum_{m=h}^{13}e^{\boldsymbol{\beta}^T\boldsymbol{x_{i,(m)}^k}}\right].
  \label{eq:mle}
  \end{aligned}
\end{equation}

Using the estimated parameters $\hat{\boldsymbol{\beta_k}}$, the VOT for run $k$ is calculated as the marginal rate of substitution between the travel cost $c$ and travel time $t$:
\begin{equation}
    VOT_k=\frac{\hat{\beta}_{time}^k}{\hat{\beta}_{cost}^k}. \label{eq:VOT}
\end{equation}

After conducting the MLE in \Cref{eq:mle} and obtaining the VOT by \Cref{eq:VOT} for each run, we use every run's factor setting and corresponding VOT to calculate the income elasticity of the VOT by conducting a linear regression, which is specified by,
\begin{equation} \label{eq:elasticity_reg}
\begin{split}
\ln(VOT)&=b_0+b_1\times\ln(\texttt{income})+b_2\times \texttt{male}+b_3\times \texttt{age50s}+b_4\times \texttt{highschool} \\
&+b_{5}\times \texttt{commute}+b_{6}\times\texttt{leisure}+b_{7}\times\texttt{personal}.
\end{split}
\end{equation}
Here, $b_1, b_2, \dots, b_{7}$ are the parameters estimated in the regression. $\ln(\text{income})$ is the natural logarithm of respondents’ income; $\texttt{male}$, $\texttt{age50s}$, and $\texttt{highschool}$ are the binary indicators for gender (reference = female), age (reference = 20s), and education level (reference = bachelor's), respectively; the dummy variables $\texttt{commute}$, $\texttt{leisure}$, and $\texttt{personal}$ indicate travel purposes, where the reference category is business travel. 
The estimated coefficient for the natural log of income, i.e. $\hat{b}_1$, is the income elasticity. 

In addition, we calculate the income elasticity of the VOT for different travel purposes separately. For the VOT estimated by each choice setting, we categorize the results into four groups based on the purpose of travel, and conduct another linear regression, 
\begin{equation} \label{eq:elasticity_sep}
\ln(VOT)=b_0+b_1\times\ln(\texttt{income})+b_2\times\texttt{male}+b_3\times\texttt{age50s}+b_4\times\texttt{highschool}.
\end{equation}
The coefficients $b_1$, $b_2$, $b_3$, and $b_4$ and the variables are the same as \Cref{eq:elasticity_reg}; similarly, the estimated coefficient for the natural log of income, i.e. $\hat{b}_1$, is the income elasticity.

\section{Results} \label{sec:results}
\subsection{The Aggregated VOT of LLMs}
\begin{figure}[h!]
\centering
\includegraphics[width=0.7\textwidth]{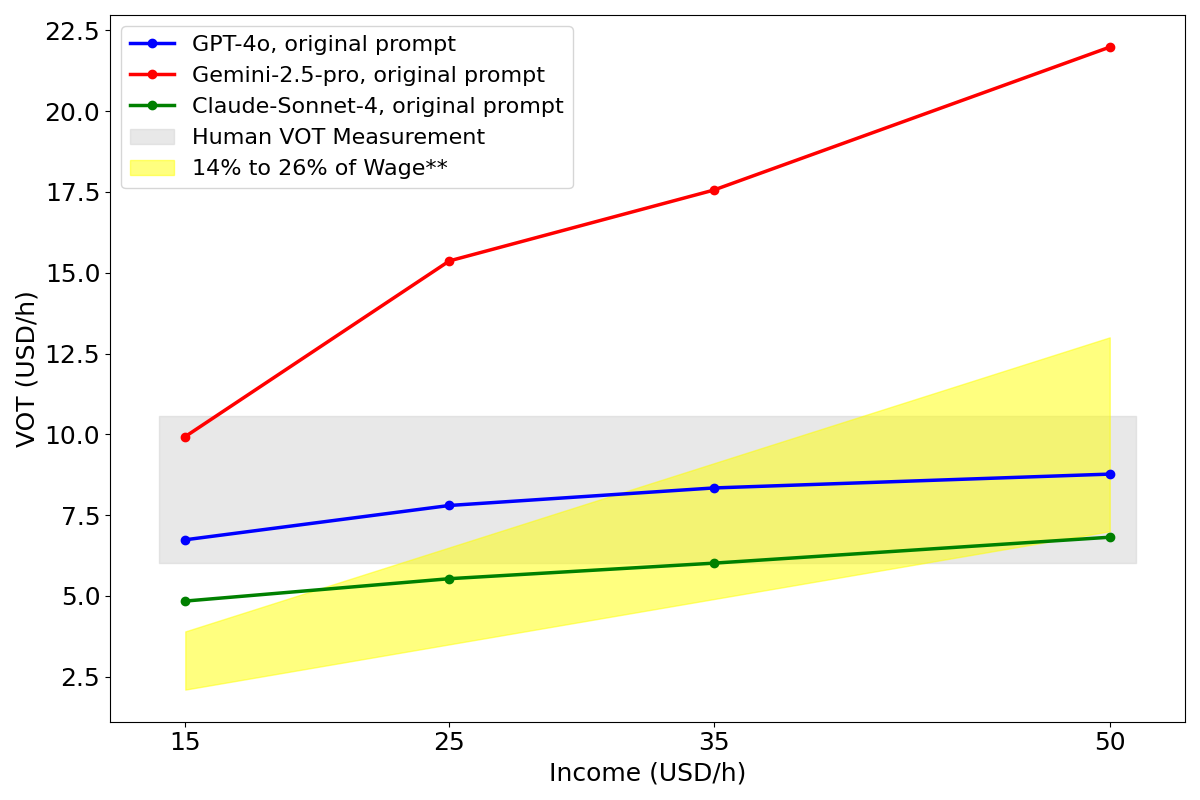}
\caption{The average VOT-income plot of LLMs, estimated with choice package 1}
\label{fig:vot_agg}
\end{figure}

\hspace{1.5em}We first present the calculated VOT of three LLMs at an aggregated level. For each income level, we use package 1 to calculate the average VOT of three models under all combinations of context factors enumerated in \Cref{tab:factor}, compare them with the results from \cite{calfee2001} in \Cref{fig:vot_agg}. 

The average VOT of GPT-4o, Claude-Sonnet-4, and  Gemini-2.5-pro are \$7.92/h, \$5.80/h, and \$14.37/h, while their average VOT-income ratios are 0.295, 0.213, and 0.489, respectively. In comparison, after accounting for inflation, the human VOT estimated by \cite{calfee2001} ranges from \$6.02/h to \$10.56/h\footnote{This value is lower than typical VOT measurement, such as the USDOT's suggested value \citep{USDOT2016VTT}. This difference may be attributed to Calfee et al.'s experiment design and does not impact our comparison. For a more detailed discussion on this matter, the readers can refer to \cite{brownstone2005valuing}.}, and the human VOT-income ratio obtained by them is between 0.14 and 0.26. For GPT-4o and Claude-Sonnet-4, their VOT exhibits a range similar to that of humans, but the VOT-income ratios of the lower-income population groups exceed those of humans, whereas the ratios for the higher-income population groups remain within the plausible range. This indicates that these two LLMs' VOT values are more aligned with higher-income groups than lower-income groups, reflecting some potential bias in the LLMs' representation of different income groups. Gemini-2.5-pro demonstrates a different pattern: its VOT-income ratio varies less across income levels, yet only the VOT for the lowest income level falls within the human range, and the VOT for higher income levels are much greater than human results. At the aggregated level, all models exhibit patterns partially similar to human data. In particular, VOT increases with income level, mirroring a critical trend observed in human travelers \citep{USDOT2016VTT}. However, we also observe discrepancies between the VOT of LLMs and humans, where the discrepancy is model-specific. This is possibly related to the distinctions in the training process of different LLMs, where the training methodology may vary, and the sources of corpus data may be from different regions or populations.

\subsection{Sensitivity to Travel Context} \label{subsec: sensitivity}
\hspace{1.5em}After estimating the absolute VOT values of the LLMs, we proceed to analyze their sensitivity to socio-demographic factors and travel purposes. First, to quantify these effects, we performed a separate linear regression on the VOT estimated under choice package 1 for all three LLMs. These analyses follow the model specified in \Cref{eq:elasticity_reg}, with the results presented in \Cref{tab:lr_models}. 

{
\renewcommand{\arraystretch}{1.2}
\fontsize{11}{11}\selectfont
\setlength{\tabcolsep}{3pt}

\begin{table}[h!]
\centering
\begin{threeparttable}
\begin{tabular}{@{}>{\raggedright\arraybackslash}p{4cm} 
                >{\raggedright\arraybackslash}p{3cm} 
                >{\raggedright\arraybackslash}p{3cm} 
                >{\raggedright\arraybackslash}p{3cm}@{}}
\toprule
\textbf{Model} & \textbf{GPT-4o} & \textbf{Gemini-2.5-pro} & \textbf{Claude-Sonnet-4} \\
\midrule
\textbf{Variables} & \multicolumn{3}{c}{Estimated coefficient (Standard error)} \\
\midrule
\textbf{Intercept} & 1.368 (0.070)*** & 0.811 (0.114)*** & 1.010 (0.077)*** \\
\textbf{ln(income)} & 0.221 (0.020)*** & 0.689 (0.032)*** & 0.286 (0.022)*** \\[4pt]

\textbf{Gender} (Ref:\ Female)&&&\\
\quad Male & 0.070 (0.017)*** & 0.089 (0.029)*** & -0.092 (0.017)** \\[4pt]

\textbf{Age} (Ref:\ 20s)&&&\\
\quad 50s & -0.020 (0.017) & -0.208 (0.029)*** & -0.051 (0.017)** \\[4pt]

\textbf{Education} (Ref:\ Bachelor’s)&&&\\
\quad High school & -0.073 (0.017)*** & -0.376 (0.029)*** & -0.328 (0.017)*** \\[4pt]

\textbf{Purpose} (Ref:\ Business)&&&\\
\quad Commuting & 0.029 (0.025) & -0.227 (0.040)*** & 0.027 (0.025) \\
\quad Leisure travel & -0.143 (0.025)*** & -0.600 (0.040)*** & -0.108 (0.025)*** \\
\quad Personal travel & -0.052 (0.025)** & -0.383 (0.040)*** & -0.109 (0.025)*** \\[6pt]

\hline
$R^{2}$ & 0.645 & 0.886 & 0.844 \\
Adjusted $R^{2}$ & 0.624 & 0.879 & 0.835 \\
Observations & 128 & 128 & 128 \\
$F(7,120)$ & 31.13*** & 132.6*** & 92.98*** \\

\bottomrule
\end{tabular}

\begin{tablenotes}[flushleft]
\footnotesize
\item \textit{Significance codes}: \,***$p<0.01$;\ **$p<0.05$;\ *$p<0.1$.
\end{tablenotes}
\end{threeparttable}
\caption{Linear regression results for GPT-4o, Gemini-2.5-pro, and Claude-Sonnet-4, estimated with choice package 1}
\label{tab:lr_models}
\end{table}
}

The results in \Cref{tab:lr_models} reveal essential similarities between the VOT of LLMs and humans. To begin with, all three LLMs have the coefficients for the natural log of income positive and statistically significant, firmly demonstrating the pattern that the LLMs' VOT rises when it is assigned a higher income, a finding that aligns with both the economic foundations of VOT \citep{USDOT2016VTT} and numerous empirical studies \citep{meta2009,fournier2021impact,binsuwadan2023income}. Meanwhile, the personal and leisure dummies consistently have negative coefficients across all LLMs, all of which are statistically significant, revealing that LLMs value time more during business travel than for personal or leisure trips, which mirrors established findings from human studies \citep{binsuwadan2023income}. Moreover, all models have negative and significant coefficients for the dummy of high school education, placing a lower VOT when assigned the lower education level, which is in accord with existing research \citep{ettema2007multitasking}. 

For other context factors, the sign or significance level may slightly vary across different LLMs. In cases where they are significant, some variables also follow expected patterns. For instance, Gemini-2.5-pro assigns a higher VOT to business travel over commuting, which echoes well-documented human behaviors \citep{USDOT2016VTT,binsuwadan2023income}. With respect to gender, all LLMs demonstrate statistically significant sensitivity, however, the sign of coefficient for male dummy is not consistent across models. While some literature show that male human travelers exhibit higher VOT than females, this pattern is not universally found in prior studies \citep{fournier2021impact}. Similarly, while Gemini-2.5-pro and Claude-Sonnet-4 assign higher VOT to age group of the 20s over the 50s, there are both studies on human travelers that support this phenomenon \citep{malokin2021millennials} and contradict it \citep{fournier2021impact}. These results may indicate that while there exists VOT heterogeneity among populations of human travelers, LLMs exhibit more fixed and inherent behavioral patterns.

\begin{figure}[h]
  \centering
  \begin{subfigure}[b]{0.31\linewidth}
    \includegraphics[width=\linewidth]{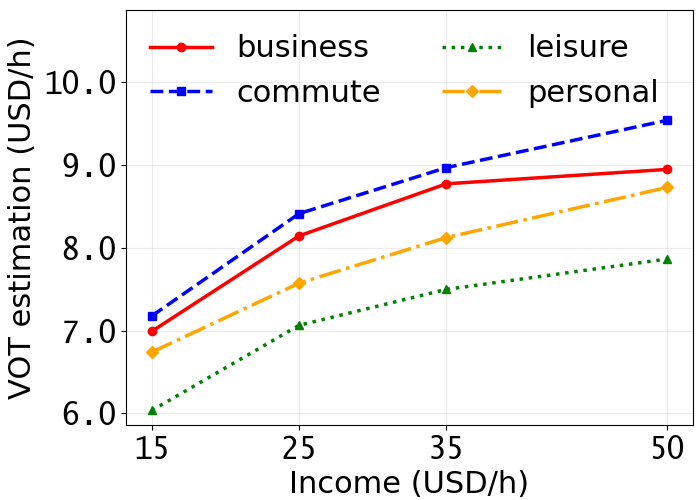}
    \caption{GPT-4o}
  \end{subfigure}\hfill
  \begin{subfigure}[b]{0.31\linewidth}
    \includegraphics[width=\linewidth]{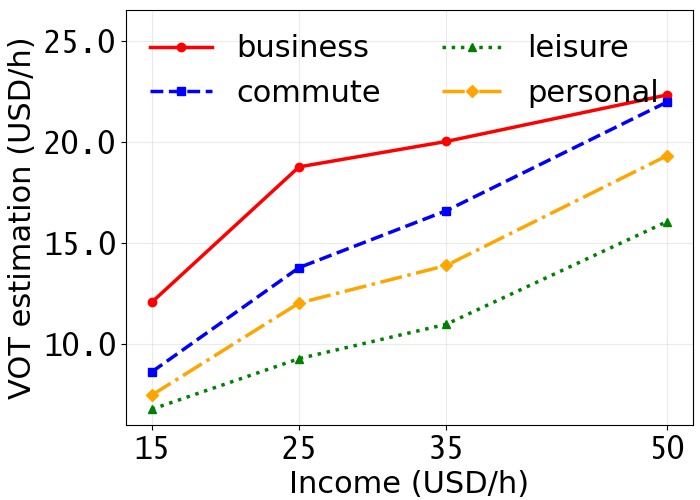}
    \caption{Gemini-2.5-pro}
    \label{fig:vot_gemini}
  \end{subfigure}\hfill
  \begin{subfigure}[b]{0.31\linewidth}
    \includegraphics[width=\linewidth]{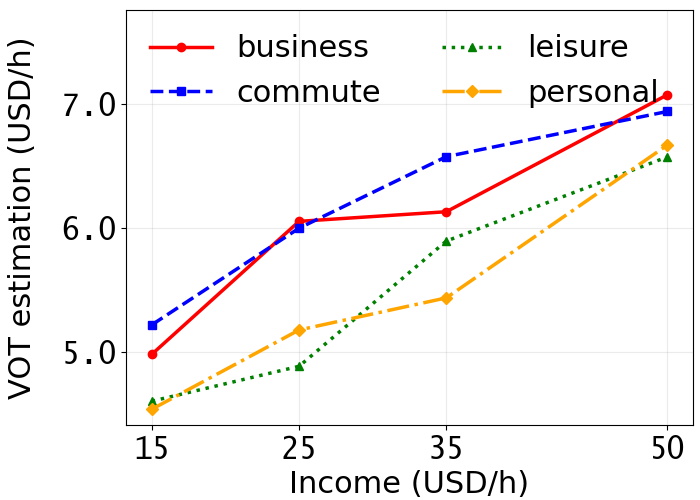}
    \caption{Claude-Sonnet-4}
    \label{fig:vot_claude}
  \end{subfigure}

  \caption{The average VOT-income plot of GPT-4o, Gemini-2.5-pro, and Claude-Sonnet-4 for different travel purposes, estimated with choice package 1}
  \label{fig:vot_models_purpose}
\end{figure}

Then, we illustrate the scale of factor impacts on the LLMs' VOT by plotting the average VOT-income curves for each factor. \Cref{fig:vot_models_purpose} presents the average VOT-income curves for different travel purposes, including the estimation results for choice package 1. For all three LLMs, the VOT exhibits a similar increasing trend for four travel purposes, with the average VOT values for commuting and business travel yielding higher VOT than leisure travel and personal travel for all income levels. This indicates that the LLMs can adjust their behavior when assigned different income levels and travel purposes, producing VOT values that are consistent with human trends. However, while we observe similar patterns that are consistent with humans in general, the results also demonstrate some model-specific discrepancies. GPT-4o yields higher VOT for commuting than for business travel, while Gemini-2.5-pro yields the opposite; meanwhile, Claude-Sonnet-4 produces similar VOT for commuting and business travel, and also produces another group of similar VOT for personal travel and leisure travel. 

\begin{figure}[p]
  \centering
  
  \textbf{\large (a) Age Group Analysis}\par\medskip
  
  \begin{subfigure}[b]{0.31\linewidth}
  \label{fig:age_models}
    \includegraphics[width=\linewidth]{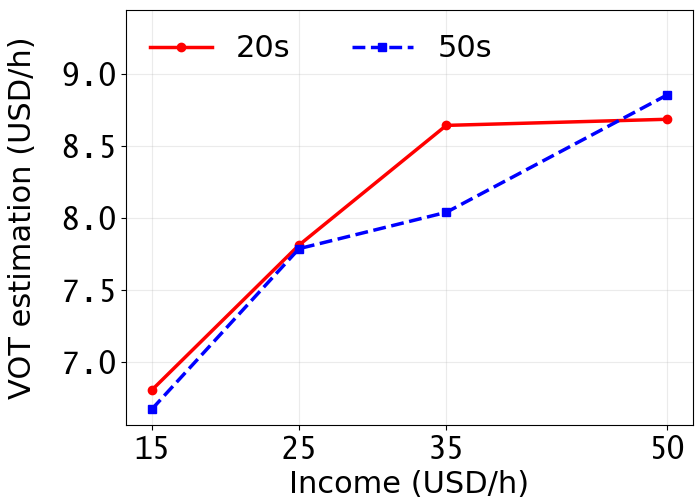}
    \caption{GPT-4o}
  \end{subfigure}\hfill
  \begin{subfigure}[b]{0.31\linewidth}
    \includegraphics[width=\linewidth]{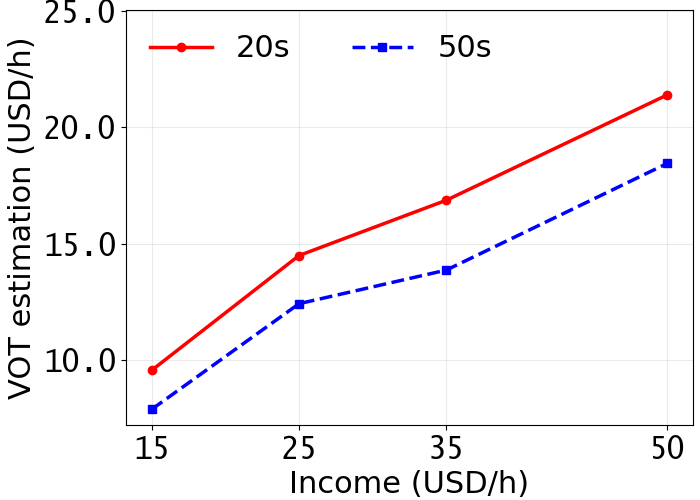}
    \caption{Gemini-2.5-pro}
    \label{fig:age_gemini}
  \end{subfigure}\hfill
  \begin{subfigure}[b]{0.31\linewidth}
    \includegraphics[width=\linewidth]{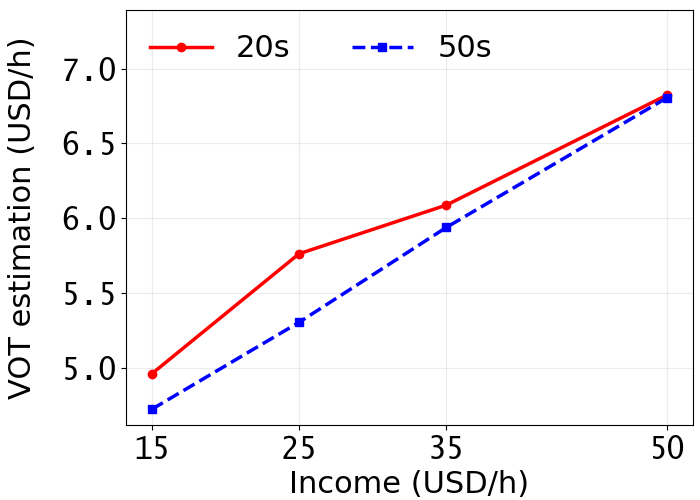}
    \caption{Claude-Sonnet-4}
    \label{fig:age_claude}
  \end{subfigure}

  \par\vspace{1cm} 

  \textbf{\large (b) Gender Group Analysis}\par\medskip

  \begin{subfigure}[b]{0.31\linewidth}
    \includegraphics[width=\linewidth]{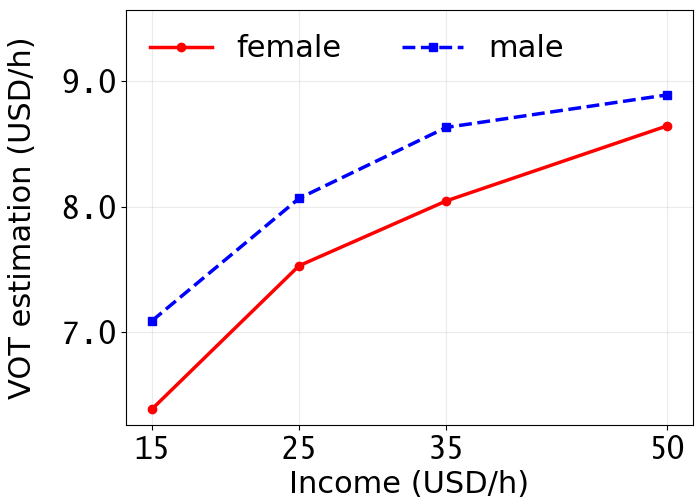}
    \caption{GPT-4o}
  \end{subfigure}\hfill
  \begin{subfigure}[b]{0.31\linewidth}
    \includegraphics[width=\linewidth]{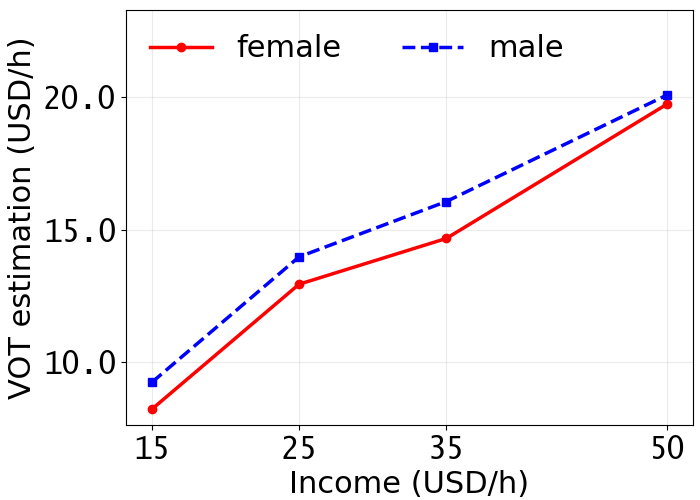}
    \caption{Gemini-2.5-pro}
    \label{fig:sex_gemini}
  \end{subfigure}\hfill
  \begin{subfigure}[b]{0.31\linewidth}
    \includegraphics[width=\linewidth]{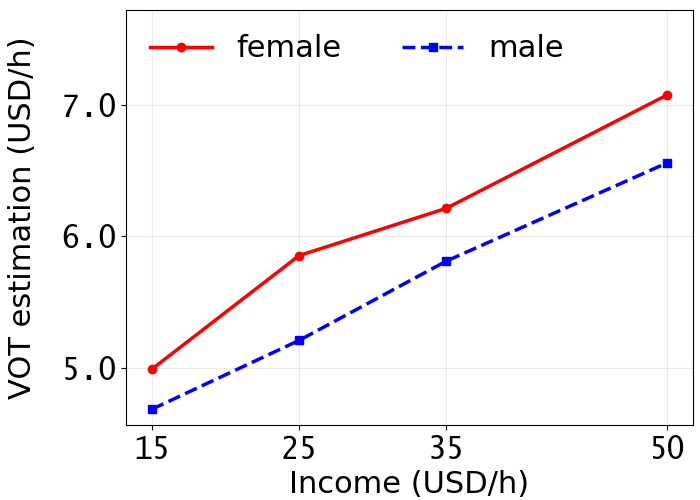}
    \caption{Claude-Sonnet-4}
    \label{fig:sex_claude}
  \end{subfigure}

  \par\vspace{1cm} 

  \textbf{\large (c) Education Group Analysis}\par\medskip

  \begin{subfigure}[b]{0.31\linewidth}
    \includegraphics[width=\linewidth]{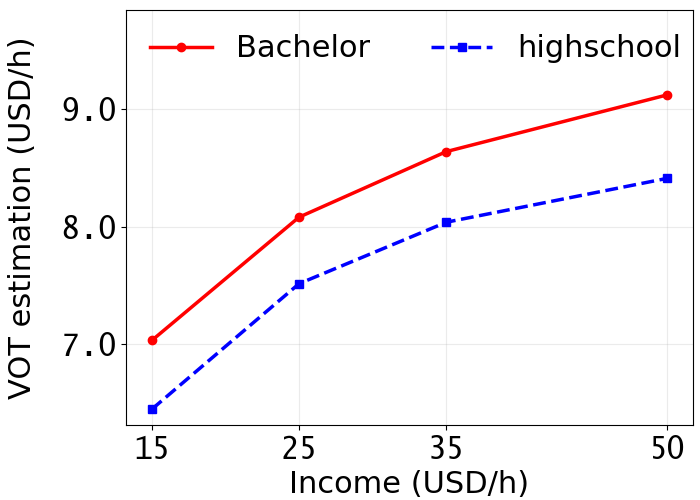}
    \caption{GPT-4o}
  \end{subfigure}\hfill
  \begin{subfigure}[b]{0.31\linewidth}
    \includegraphics[width=\linewidth]{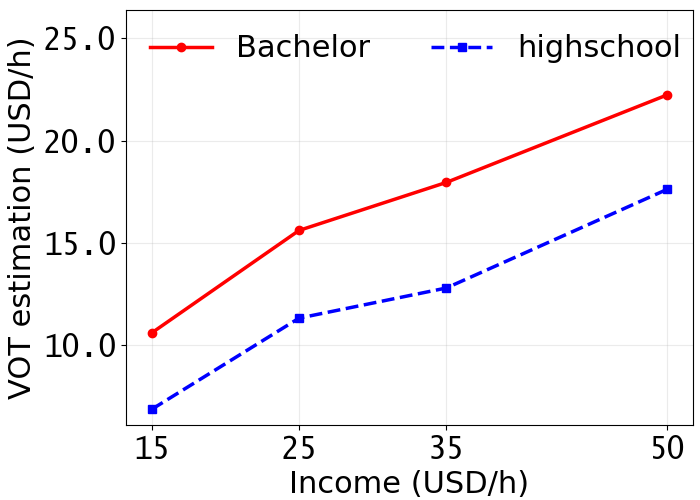}
    \caption{Gemini-2.5-pro}
    \label{fig:edu_gemini}
  \end{subfigure}\hfill
  \begin{subfigure}[b]{0.31\linewidth}
    \includegraphics[width=\linewidth]{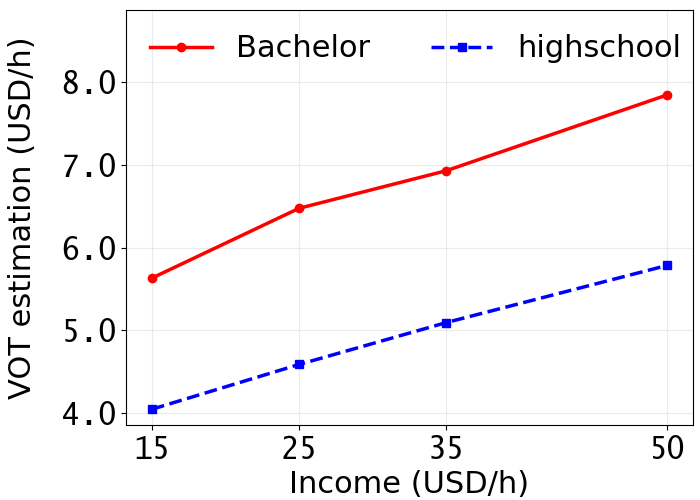}
    \caption{Claude-Sonnet-4}
    \label{fig:edu_claude}
  \end{subfigure}
  
  \caption{Value of Time (VOT) estimates across different demographic segments. Rows represent (a) Age, (b) Gender, and (c) Education groups. Columns represent the estimation models used.}
  \label{fig:combined_demographics_models}
\end{figure}

We also illustrate how the magnitude of the VOT of LLMs varies across different socio-demographic profiles. The average VOT-income curves for different ages, genders, and educational backgrounds are plotted in \Cref{fig:combined_demographics_models} a-c, d-f, g-i, respectively. The monotonically increasing pattern holds for all VOT-income plots, demonstrating that this pattern is consistent for all LLMs.
Furthermore, the plot reveals small but clear VOT distinctions by age and gender that vary by models, and a more apparent distinction by education level across all LLMs, where the VOT of higher education level is consistently higher. 
These results provide more detailed illustrations of the findings in \Cref{tab:lr_models}, demonstrating that LLMs' sensitivities to travel contexts are significant and robust. However, the VOT model-variant distinctions by gender are also consistent for each LLM respectively, indicating the possibility of LLMs inherently exhibiting model-specific bias.

\begin{figure}[h!]
\centering
\includegraphics[width=0.7\textwidth]{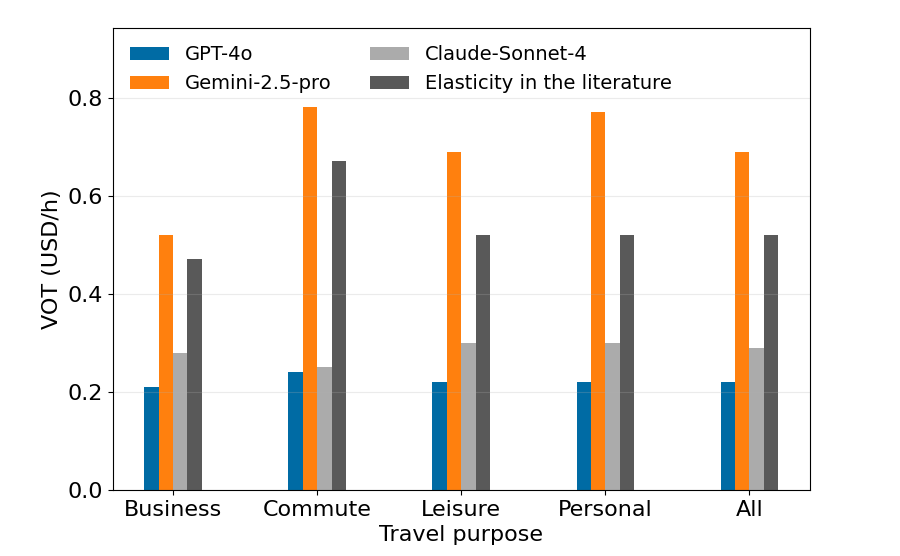}
\caption{The income elasticity of the VOT of GPT-4o, Gemini-2.5-pro, and Claude-Sonnet-4 for different travel purposes}
\label{fig:elas_models}
\end{figure}

Next, we calculate the income elasticity of VOT to examine the magnitude of LLMs' sensitivity to income and travel purposes with a critical quantitative metric. The income elasticities for the VOT of all LLMs are presented in \Cref{fig:elas_models}. For the VOT of GPT-4o and Claude-Sonnet-4, we observe that all income elasticities are below 0.3, a level significantly lower than that observed in human travelers \citep{meta2009}, and the elasticities for all travel purposes are notably close, which stands in sharp contrast with the varied elasticities of humans for business travel, commuting, and other travel purposes. In contrast, the elasticities of Gemini-2.5-pro's VOT demonstrate a similar pattern to those of humans, with the elasticity of commuting being the highest and the elasticity of business travel being the lowest, but all elasticities are significantly higher than human results. These results demonstrate clear discrepancies between the VOT of LLMs and human travelers.

Overall, the context sensitivity analysis demonstrates that LLMs' sensitivities to income, travel purpose, and other factors are consistent across models and largely parallel established human patterns, indicating that LLMs demonstrate encouraging alignment with human travelers at the behavioral value level. However, the income elasticity suggests that LLMs, albeit processing the same direction of trends and aggregate values as humans in terms of VOT, the VOT similarities between LLMs and human travelers are limited in terms of the magnitude of sensitivities. Without the lived experience of financial constraints or opportunity costs, the LLMs correctly process the direction of human behavior but fail to capture the nuanced trends under varied contexts. Such an issue needs to be accounted for in utilizing LLMs in policy simulations, particularly for economic cost-benefit analysis.

\subsection{Sensitivity to Choice Setting}

\begin{figure}[h!]
\centering
\includegraphics[width=0.53\textwidth]{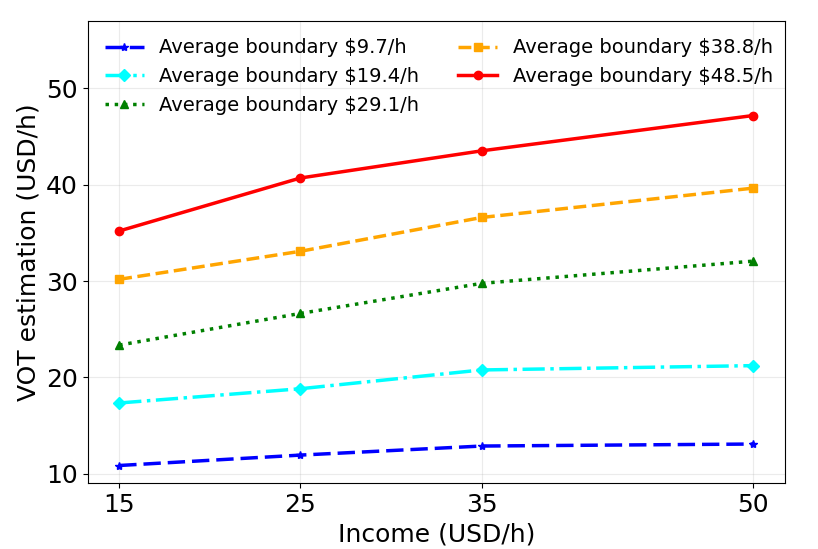}
\caption{The VOT-income plot for the estimations of GPT-4o using choice sets with different trade-off ratios}
\label{fig:vot_varset}
\end{figure}

\hspace{1.5em}
While the previous section establishes a baseline comparison across multiple models, this section shifts focus to a deep-dive analysis of sensitivity to choice settings. To manage the computational complexity of the full factorial design involving multiple trade-off ratios, we utilize GPT-4o as the representative model for this extensive granular analysis. We first illustrate the scale of impact from choice setting on the LLM's VOT by plotting the average VOT-income curves for each choice package. The average VOT-income plot under each trade-off ratio is illustrated in \Cref{fig:vot_varset}. The results show that for each choice setting, the LLM's VOT exhibits a clear increasing trend with income, which aligns with findings for human VOT. Furthermore, controlling for income level, as the average trade-off ratio increases, the average VOT increases correspondingly, indicating that the trade-off ratios of the choice package positively influence the VOT of the LLM. This is consistent with the study of human VOT \citep{hess2018}, highlighting deeper similarities between the LLM and humans in terms of sensitivity to the choice setting. On the scale of the trade-offs' impact, for human subjects, \cite{hess2018} found that doubling the trade-off level increases VOT by 60-233\%, while quintupling it yields a 198-340\% increase. As shown in \Cref{fig:vot_varset}, the magnitude of the LLM’s response consistently falls within these human ranges across all income levels, implying a deeper similarity to humans in its decision-making process.

Next, to verify the robustness of the LLM's sensitivity, we perform a separate linear regression for the VOT estimated under each choice package. These analyses follow the model specified in \Cref{eq:elasticity_reg}, with the results presented in \Cref{tab:lr}. 

{
\renewcommand{\arraystretch}{1.2}

\fontsize{11}{11}\selectfont
\setlength{\tabcolsep}{3pt}
\begin{table}[t!]
\centering
\begin{threeparttable}
\begin{tabular}{@{}>{\raggedright\arraybackslash}p{4cm} llllll@{}}
\toprule
\textbf{Trade-off ratio} 
& \$6.6/h & \$9.7/h & \$19.4/h & \$29.1/h & \$38.8/h & \$48.5/h
\\
\midrule
\textbf{Variables} & \multicolumn{6}{c}{\makecell{Estimated  coefficient\\(Standard error)}} \\
\midrule
\textbf{Intercept}                    & \makecell[l]{1.368***\\(0.070)} & \makecell[l]{1.861***\\(0.069)} & \makecell[l]{2.439***\\(0.065)} & \makecell[l]{2.450***\\(0.059)} & \makecell[l]{2.795***\\(0.059)} & \makecell[l]{2.978***\\(0.055)} \\
\textbf{\makecell[l]{Natural log of\\income}}        & \makecell[l]{0.221***\\(0.020)} & \makecell[l]{0.164***\\(0.019)} & \makecell[l]{0.175***\\(0.018)} & \makecell[l]{0.269***\\(0.017)} & \makecell[l]{0.228***\\(0.016)} & \makecell[l]{0.243***\\(0.015)} \\
\textbf{Gender} (Ref:\ Female)&&&\\
\quad Male                            & \makecell[l]{0.070***\\(0.017)} & \makecell[l]{0.065***\\(0.017)} & \makecell[l]{0.055***\\(0.016)} & \makecell[l]{0.058***\\(0.015)} & \makecell[l]{0.045***\\(0.015)} & \makecell[l]{0.034**\\(0.014)} \\
\textbf{Age} (Ref:\ 20s)&&&\\
\quad 50s                             & \makecell[l]{-0.020\\(0.017)}  & \makecell[l]{-0.027\\(0.017)}  & \makecell[l]{0.008\\(0.016)}    & \makecell[l]{0.060***\\(0.015)} & \makecell[l]{0.039**\\(0.015)} & \makecell[l]{0.037***\\(0.014)} \\[4pt]

\textbf{Education} (Ref:\ Bachelor’s)&&&\\
\quad High school                     & \makecell[l]{-0.073***\\(0.017)} & \makecell[l]{-0.026\\(0.017)} & \makecell[l]{-0.019\\(0.016)} & \makecell[l]{-0.058***\\(0.015)} & \makecell[l]{-0.032**\\(0.014)} & \makecell[l]{-0.062***\\(0.014)} \\[4pt]

\textbf{Purpose} (Ref:\ Business)&&&\\
\quad Commuting                       & \makecell[l]{0.029\\(0.025)}     & \makecell[l]{0.013\\(0.024)}   & \makecell[l]{-0.064***\\(0.023)} & \makecell[l]{-0.066***\\(0.021)} & \makecell[l]{0.001\\(0.020)} & \makecell[l]{-0.078***\\(0.019)} \\
\quad Leisure travel                  & \makecell[l]{-0.143***\\(0.025)} & \makecell[l]{-0.109***\\(0.024)} & \makecell[l]{-0.156***\\(0.023)} & \makecell[l]{-0.130***\\(0.021)} & \makecell[l]{-0.112***\\(0.020)} & \makecell[l]{-0.115***\\(0.019)} \\
\quad Personal travel                 & \makecell[l]{-0.052**\\(0.025)}   & \makecell[l]{-0.081***\\(0.024)} & \makecell[l]{-0.116***\\(0.023)} & \makecell[l]{-0.059***\\(0.021)} & \makecell[l]{-0.063***\\(0.020)} & \makecell[l]{-0.114***\\(0.019)} \\[6pt]
\hline
$R^{2}$ & 0.645 & 0.513 & 0.566 & 0.744 & 0.693 & 0.732 \\
Adjusted $R^{2}$ & 0.624 & 0.485 & 0.541 & 0.729 & 0.675 & 0.716 \\
Observations & 128 & 128 & 128 & 128 & 128 & 128 \\
$F(7,120)$ & 31.13*** & 18.06*** & 22.36*** & 49.71*** & 38.74*** & 46.84*** \\

\bottomrule
\end{tabular}
\begin{tablenotes}[flushleft]
\footnotesize
\item \textit{Significance codes}: \,***$p<0.01$;\ **$p<0.05$;\ *$p<0.1$.
\end{tablenotes}
\end{threeparttable}
\caption{Linear‐regression results of GPT-4o for different choice packages}
\label{tab:lr}
\end{table}
}

In the results presented in \Cref{tab:lr}, the consistency of LLM behavior across different contexts can be assessed by the consistency of the sign and significance of different socio-demographic and context variables across trade-off levels. Overall, the results demonstrate that the LLM's sensitivity socio-demographic and contextual factors holds across multiple choice settings. Similar to the results in \Cref{subsec: sensitivity}, the impact on VOT of three vital variables: log-income, and the personal travel and leisure travel dummies, remains statistically significant, and each of them maintains the same sign across all income levels. This suggests that the LLM maintains a consistent identity and response to travel purposes even when the trade-off level changes, revealing a high level of its internal decision-making stability. 
For gender, the coefficient is consistently positive and statistically significant. For the other variables, while the impact does not stay statistically significant across all trade-off levels, those significant parameters retain the same sign for each variable. For instance, the coefficients for the older age group dummy and commuting dummy variables are significant only in a subset of the estimations, which suggest that the GPT-4o's VOT does not differentiate significantly between commuting and business travel or between age groups. The significance of coefficients reveal that GPT-4o is clearly sensitive to some critical factors in the travel context in a highly consistent manner, while the sensitivities to some other factors are also robust yet apparently weaker.

Then, we compare the trend of the LLM's behavior with that of humans by examining the sign of the estimated parameters. The positive coefficients for log-income indicate that the LLM's VOT rises when it is assigned a higher income, which is consistent with the previous results in \Cref{subsec: sensitivity}, and also consistent with human VOT \citep{USDOT2016VTT,meta2009,fournier2021impact,binsuwadan2023income}. In contrast, the negative coefficients for the personal and leisure dummies reveal that the model values time more during business travel than for personal or leisure trips, which again accords with both previous results and the patterns of human travelers \citep{binsuwadan2023income}. These results not only demonstrate GPT-4o's consistency on exhibiting critical human-like behaviors, but also highlight the robustness of behavioral similarities to human travelers across multiple LLMs. On the impact of gender, the model consistently exhibits a slightly higher VOT when assigned male profile than female. Since there exist studies both supporting \citep{pourhashem2022gender} and not supporting this trend \citep{fournier2021impact}, this may be a robust sensitivity only representative of some populations. The coefficients of other variables may not be significant across all estimations, but for where they remain significant, the patterns still consistent. For instance, the negative coefficients for the high school dummy implies that the LLM, like humans, places a higher value on travel time when assigned a higher education level; similarly, for the significant cases, the VOT of business trips are higher than commuters. These results follow the similar patterns with both human VOT \citep{ettema2007multitasking,USDOT2016VTT,binsuwadan2023income,malokin2021millennials} and the results observed on other two LLMs (presented in \Cref{tab:lr_models}. Regarding the age impact, GPT-4o yields higher VOT values to the older age group than to the younger age group. We observe the opposite results from other two LLMs (also presented in \Cref{tab:lr_models}), while human studies demonstrate diverse conclusions on which age group exhibits higher VOT values. This reveals that the discrepancies across LLMs regarding the sensitivities to certain factors are intrinsic, signifying that different LLMs may have better behavioral representations of different populations; however, within this same model, the stability observed on GPT-4o, even among less consistently significant variables, further reinforces the model's non-arbitrary and principled response to changing choice contexts, demonstrating that the LLM's behaviors are highly robust. 


\begin{figure}[h!]
\centering
\includegraphics[width=0.92\textwidth]{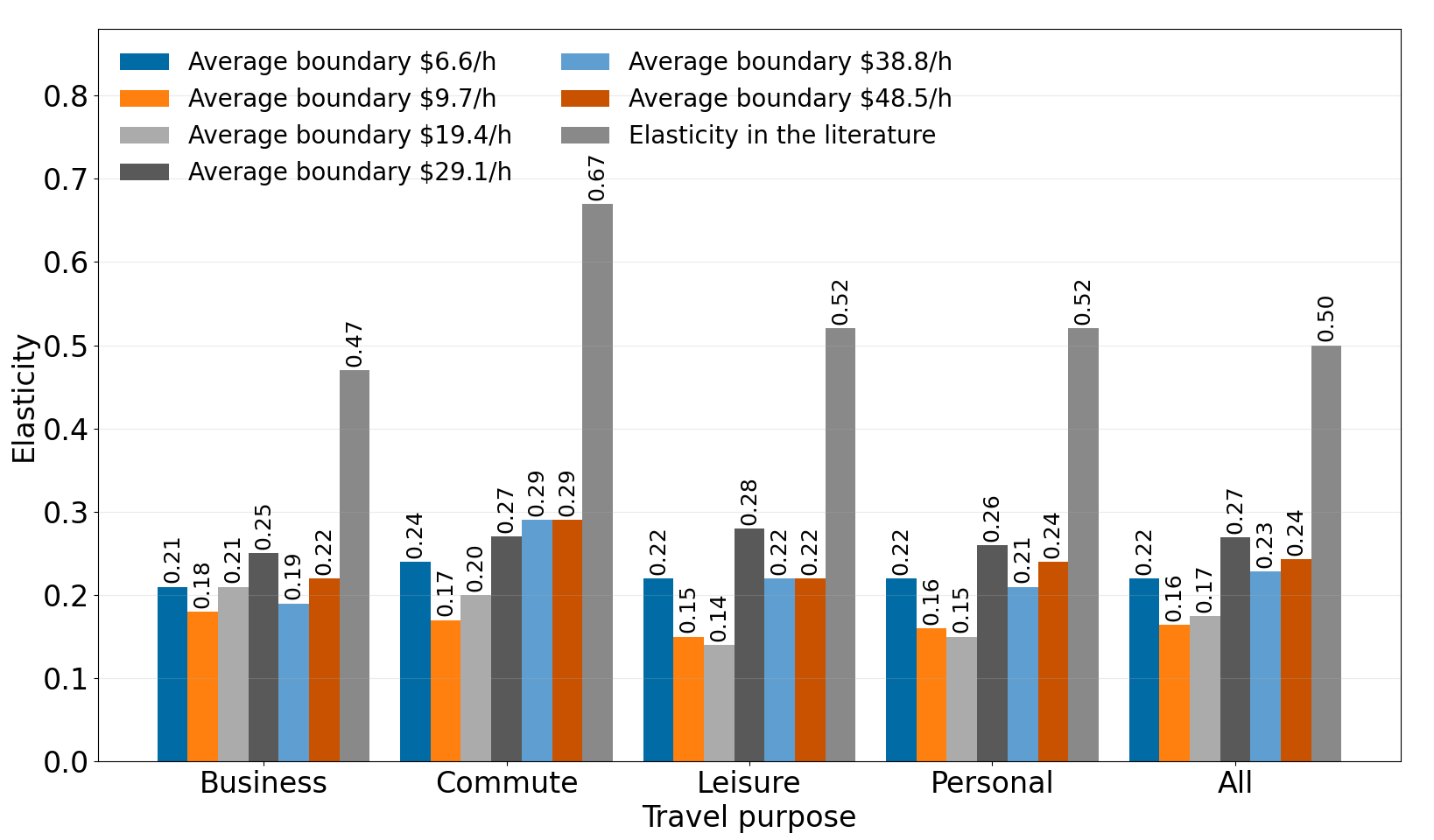}
\caption{The income elasticity for different travel purposes, estimated using choice packages with different trade-off ratios}
\label{fig:elas_varset}
\end{figure}

Finally, we calculate the income elasticity of VOT under different choice settings, where the results are presented in \Cref{fig:elas_varset}. 
We observe that all income elasticities of the LLM's VOT are below 0.3, a level significantly lower than that observed in human travelers \citep{meta2009}. This suggests that the GPT-4o's behavior is less sensitive to changes in income than humans. Moreover, the elasticities for all travel purposes are notably close, which stands in sharp contrast with the varied elasticities of humans for business travel, commuting, and other travel purposes. This indicates that, while being consistent in value, GPT-4o exhibits less income elasticity compared to humans across different travel purposes, suggesting that GPT-4o has less behavioral sensitivity compared to its human counterpart. Since this characteristic is not observed in all LLMs, it may indicate that model-specific validation and behavioral alignment is necessary before an LLM is utilized for downstream applications.

In summary, this section provides additional insights in GPT-4o's behavioral fidelity. The model demonstrates high internal stability, maintaining consistent preference structures across varied choice packages. At the same time, issues of lower income elasticity also persist across packages. This implies that calibration is essential regardless of choice package design to have the model accurately reflect the magnitude of human economic trade-offs.

\section{Conclusion} \label{sec:conclusion}

\hspace{1.5em}In this study, we present a novel analysis of LLMs' travel behavior through the lens of VOT. Our approach treats three popular LLMs, GPT-4o, Gemini-2.5-pro, and Claude-Sonnet-4, as survey respondents, allowing us to systematically analyze their responses using established methods from travel behavior research. We devise a comprehensive full factorial experiment that assigns LLMs distinct socio-demographic personas and places them in various travel contexts, including different trip purposes and choice settings with varying time-cost trade-offs, and quantify their impact on LLMs' VOT using rank-ordered logit and linear regression models. We then evaluate the LLMs' decision-making consistency and compare their values and sensitivities with those of humans. Our main findings on the VOT of LLMs and the implications are summarized below.

\begin{itemize}
    \item The VOT values exhibited by LLMs are model-specific. While GPT-4o and Claude-Sonnet-4 demonstrate VOT ranges that are similar to human data (albeit closer to high-income groups), the VOT of Gemini-2.5-pro significantly exceeds typical human benchmarks. These observations provide positive evidence for the behavioral alignment between some LLMs and human travelers, signifying the potential of these LLMs in applications such as synthetic data generation. However, this also necessitates the selection of suitable models and careful validations of LLMs' behaviors before these applications.


    \item LLMs demonstrate consistent sensitivity to travel contexts. The VOT values of all three LLMs show a stable increasing pattern with income and exhibit higher VOT for business travel and commuting, which aligns well with observations on human travelers. Furthermore, the impacts from travel purpose and income on all three LLMs maintain the same direction and statistical significance under varied choice settings. This consistent, human-aligned behavior across different contexts implies that LLMs' decision-making is coherent rather than random, suggesting their potential as a proxy for simulating travel preferences.

    \item The LLMs' behavior changes in a human-like manner in response to different choice settings. Exploring GPT-4o in depth, we discover that the trade-off ratios of time and cost in the choice setting have a clear positive impact on the LLM's VOT. This finding reveals another critical aspect of sensitivity that is consistent with humans, suggesting the model's alignment with humans extends beyond surface-level socio-demographic factors to more fundamental principles of decision-making. Furthermore, it carries a practical implication. Researchers could leverage such LLMs to generate initial hypotheses for novel scenarios, particularly those where collecting human data is costly, difficult, or otherwise impractical.

    \item One notable limitation is that the context sensitivities of the LLMs diverge significantly from humans in specific dimensions, particularly income elasticity, which is consistently different than humans across all travel purposes. This discrepancy indicates that the current model may not be able to capture the economic sensitivity to the incomes of travelers. Therefore, further calibration of LLMs is critical to address this issue and achieve the behavioral fidelity necessary for the LLMs' reliable application in transportation planning.

    \item Another critical divergence lies in the non-uniformity of human representation across different models. Our analysis reveals that different LLMs may exhibit opposite sensitivities to certain contextual factors. Furthermore, these model-specific behaviors sometimes align with conflicting subgroups of human empirical literature, suggesting that no single LLM captures the full heterogeneity of the human population. This phenomenon underscores that LLMs are not universal proxies for all human travelers. Therefore, an LLM's behavioral characteristics must be validated against the target population before its deployment.
    
\end{itemize}

Overall, our findings demonstrate LLMs' capacity to comprehend travel context and make human-like travel decisions. Our analysis indicates that LLMs provide a non-perfect but valuable basis for simulating human travel behavior. The qualitative alignment between LLMs and human travelers and the convenience of LLM-subject experiments have practical implications, creating a tool for applications such as preliminary screening, hypothesis generation, and synthetic data production. Researchers and practitioners can leverage LLMs to gain initial insights into novel scenarios or conduct indicative analysis under previously unexamined situations where human data are scarce. However, due to the model-specific elasticity and human representation of LLMs' output, we also recommend using unmodified LLMs with caution for quantitative assessments and conducting validation in advance, as existing bias or specific demographic idiosyncrasy may compromise the validity of the analysis and mislead public policy. Furthermore, choosing suitable models based on their specific behavioral profiles and conducting additional conditioning of the models are critical before their deployment.

While this study provides a foundational analysis, its deliberate focus on a specific context naturally presents limitations and opens up promising avenues for future research. Our investigation centers on three close-sourced LLMs, contextualized with human data exclusively from the United States. This focused approach invites a broader comparative study; future research could extend our methodology to a diverse portfolio of LLMs, including leading open-source alternatives and foundation models specifically tuned for economic or transport domains.  Furthermore, expanding the baseline to include human VOT data from various international regions would be essential to test the cross-cultural representativeness and generalizability of these models' behaviors. Methodologically, our use of zero-shot prompting was designed to probe the model's baseline capabilities without intervention. A crucial next step would be to investigate whether the model's behavioral fidelity can be actively improved. Future work could explore various alignment techniques, from few-shot prompting with contextual examples to fine-tuning on domain-specific datasets, to determine the most effective means of enhancing LLMs' value alignment for real-world applications. Additionally, our analysis is centered on the economic construct of VOT. Future work could explore other rich dimensions of travel behavior, such as risk perception, multitasking abilities, or the influence of social factors, to build a more holistic behavioral profile of these AI agents. Finally, we acknowledge that the human benchmark utilized \citep{calfee2001} may predate recent shifts in travel habits. While we explicitly adjusted monetary values accounting for inflation, the evolution of behavior remains a potential constraint on the direct generalization of these specific comparative baselines.


\section*{Declaration of Generative AI and AI-assisted technologies in the writing process}
\hspace{1.5em}During the preparation of this work, the author(s) used Google Gemini in order to conduct language editing. After using this tool/service, the author(s) reviewed and edited the content as needed and take(s) full responsibility for the content of the publication.

\bibliographystyle{apalike} 
\bibliography{mybib}

\end{document}